# Efficient Editor Generation for Compositional DSLs in Eclipse

Holger Krahn, Bernhard Rumpe, and Steven Völkel

Institute for Software Systems Engineering
Niedersächsische Technische Hochschule, Standort Braunschweig, Germany
`http://www.sse-tubs.de`

**Abstract.** When a domain specific language (DSL) shall be commonly used, it is important for the development team to have a comfortable editor well integrated in the project's development environment. As DSL are rather often subject to changes, efficient development and evolution of a comfortable editor is an important success criterion for the acceptance of domain specific languages. In this paper we demonstrate how this issue is addressed in the MontiCore DSL development framework. Basically an extension of the MontiCore DSL definition language can be used to efficiently generate DSL editors for Eclipse. The generation tool also supports the compositional language definition features of MontiCore and allows therefore the reuse of existing language and editor definitions.

## 1 Introduction

Domain specific languages (DSL) have a crucial advantage over general purpose languages. They allow us to describe the elements, relations and constraints of a certain domain a lot more concise and compact. However, there are many domains and technical spaces out there and we will see a lot of new DSLs coming up in the future, many of them to be used in software development projects. Especially these types of DSLs are predestinated for a textual realization [4].

The development of a new, but usable domain specific language involves various activities. Far too often the development activities for a DSL usually focus on concrete and abstract syntax only. Sometimes also an informal document, describing the semantics is provided, and of course the development of appropriate tooling to assist the purpose the DSL is a central aspect. This usually involves a code generator to transform the language into an executable form, but sometimes also test code generation or simulation to study essential properties of the system under development are needed.

The editor environment for such a newly defined language is a far too often neglected labor intensive development work that does not directly contribute to the project, but only helps the developers to become more effective later on. Therefore, a comfortable editor is an important success factor for a DSL, when a new, and for other developers unknown DSL has to compete with a general purpose language (GPL) like Java. For GPLs the existing tools are usually mature and support the user with a sophisticated user interface. DSLs therefore must be supplemented by a similar development environment.

The Eclipse platform [1] is an ideal target for this kind of approach. First, the platform supports the user by a full-functional Java-IDE including among others an incremental compiler. Second, the user is supported by various build and version management utilities which are essential for efficient software development. Third, due to Eclipse's extensible nature, it allows integrating new plugins for DSLs. The downside of this approach is, that due to the flexibility and power of the Eclipse platform, the provided interfaces for DSL-tooling are rather complex and still lack stability. This means a lot of experience is needed to develop and maintain a sophisticated language support. To make our MontiCore DSL framework [2, 3, 8–10, 14] as introduced in Section 2 more comfortable for language development, we have added technology to the framework that helps deriving DSL editor configurations based on their language definition.

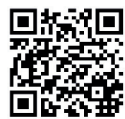

For this purpose we identified an extension of the MontiCore language definition format [10] that allows the generation of specific Eclipse plugins to integrate editors and code generators in the Eclipse platform without further knowledge about the platform itself. Therefore the developer can concentrate on the language development and the tool integration is developed and maintained inside the MontiCore framework for every DSL.

The rest of this paper is structured as follows. In Section 2 we explain the basic principles of our extended grammar format that is used to define languages in the MontiCore framework. In Section 3 we explain how an extension of this grammar format can be used to generate Eclipse plugins. To explain this extension a bit further, we applied it to a demonstrating example in Section 4. Section 5 explains other approaches whereas Section 6 concludes this paper.

## 2 Language definition using MontiCore

Using the MontiCore framework a DSL is defined by an extended context-free grammar that determines concrete textual syntax as well as the internal representation (abstract syntax). The grammar is similar to the input language of common parser generators, but allows defining non-compositional associations and inheritance for the abstract syntax. Fig. 1 contains a simple demonstrating example with three nonterminals Greeting, Person, and VIP. Further details can be found in [10].

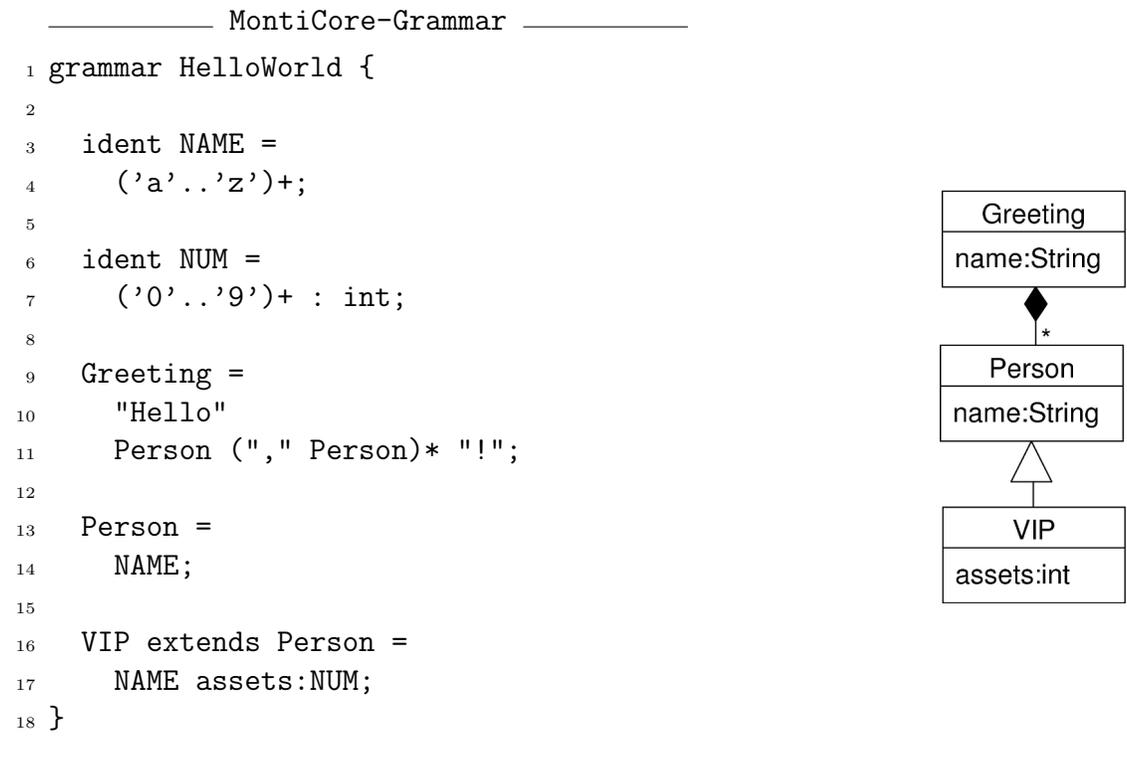

```
       ———————— MontiCore-Grammar ————————
 1 grammar HelloWorld {
 2
 3   ident NAME =
 4     ('a'..'z')+;
 5
 6   ident NUM =
 7     ('0'..'9')+ : int;
 8
 9   Greeting =
10     "Hello"
11     Person ("," Person)* "!";
12
13   Person =
14     NAME;
15
16   VIP extends Person =
17     NAME assets:NUM;
18 }
```

**Fig. 1.** Example of an MontiCore grammar and induced abstract syntax

A production has a name and body (right hand side) separated by "=". The body contains nonterminals (reference to other rules) and terminals. The usual concepts for structuring the body are alternatives (separated by "|") and blocks (in parenthesis). Blocks, nonterminals and terminals can have the cardinality "?" (optional), "*" (unbounded cardinality) or "+" (at least one). Furthermore, blocks, nonterminals and terminals respectively their presence can be named (in form of a prefix like "assets:") in order to access these elements in the abstract syntax.

The abstract syntax is realized by heterogeneous typed Java classes where the named elements are represented as attributes. To distinguish them from further attributes that can be added to mimic attribute grammars [7], we call these special form of synthesized attributes *production attributes*.

The MontiCore grammar supports two kinds of modularity concepts: First, multiple language inheritance (similar to [12]) to reuse a language definition by specifying the delta only. Second, language embedding where language fragments with explicit holes are used. These holes have to be filled by other grammars at configuration time to form a complete language.

Multiple languages can be bundled to a *DSLTool* that executes algorithms, so called *workflows*, on the languages instances. The framework simplifies reoccurring challenges like error messages, file creation, loading of depending instances and logging. The MontiCore grammar language has a flexible *concept* definition mechanism to specify further properties of the language under design. In the following the concept named *texteditor* is used to define the Eclipse integration.

## 3 Generation of Editors

We divide the generation of Eclipse-based editors into three levels. First, the *fragment level* contains functionalities which depend on language fragments (including language inheritance) as described in Section 2. Second, functionalities which are specific for a complete language or depend on the composition of fragments to a complete language are defined on the *language level*. And third, the *tool level* describes functionalities which depend on several languages bundled to a deployable Eclipse plugin.

### 3.1 Fragment Level

Grammar fragments can be enriched by editor attributes which are used to define several aspects of a generated texteditor specific for the current grammar fragment. In order to define these aspects, the concept called *texteditor* is included in a fragment definition whose functionalities are introduced in the following.

**Syntaxhighlighting.** MontiCore offers a build-in mechanism for the usage of comments similar to Java: single-line comments are preceded by `//` and are terminated at the end of the line whereas multi-line comments are delimited by `/*` and `*/`. These comment definitions are also used in the generated editor: Each language instance is partitioned into comment (in green) and code regions. Furthermore, keywords which will be colored in magenta inside code regions are defined by a comma-separated list.

**Folding.** Folding allows to collapse/uncollapse code regions in order to get more overview of the language instance. Small symbols are shown next to different code regions, a minus as symbol to collapse a region, a plus as symbol to uncollapse. When collapsing a region, the according code is abbreviated by the first line followed by two dots. Doing so, the delevoper can concentrate on the important parts whereas the uninteresting but necessary parts (e.g., the list of imports in Java) can be hidden.

MontiCore connects the definitions of nonterminals and the definitions of regions to be foldable. A comma-separated list of nonterminal names is used to define which regions can be folded in the generated editor.

**Outline.** An outline visualizes the structure of a language instance in a tree view which consists of *segments* representing different parts of the instance. MontiCore allows to define nonterminal-specific segments. Therefore, each segment is related to a nonterminal and can include a path to an icon which can be used in order to visualize the type of the segment and an instance-specific text which consists of combinations of static text and production attributes of the corresponding nonterminal. Figure 2 shows an addition to the grammar from Figure 1. In this example, each instance of the nonterminal `Person` is represented by a segment in the outline containing the defined icon and the word "person" followed by his name. Note that it is possible to use an arbitrary combination of static strings and production attributes.

```
                        MontiCore-Grammar
1 concept texteditor {
2     segment: Person ("pict/person.gif") show: "Person" name;
3 }
```

**Fig. 2.** Definition of segments in MontiCore

**Compositionality on Fragment Level.** The compositionality on the fragment level is particularly targeted at grammar inheritance which is not only reflected in the abstract and concrete syntax of the involved languages but also in the functionality of the generated editors. Editor attributes defined for a supergrammar are reused in order to generate editors for a subgrammar. Therefore, all keywords, foldings, and outline segments of the supergrammar are automatically evaluated whereas it is still possible to customize them. Doing so, it is possible to extend the list of keywords by defining the additional keywords only. Furthermore, it is possible to add new foldings and segments by defining them in the normal way.

By this strategy, it is possible to create a library of language fragments including their editor definitions which then can serve as basis when defining a new language by inheritance or embedding. This minimizes the effort for developing new DSLs as well as their specific editors.

### 3.2 Language Level

Several functionalities are dedicated to a concrete combination of fragments in a specific editor and therefore, cannot be expressed at the fragment level. A prominent example are error messages: compiler errors and errors occurring in a syntactic analysis (also denoted by semantic analysis in compiler theory) typically depend on which fragments are combined. Furthermore, editor actions, such as code formatting, code generation, or model composition are often developed for a specific fragment combination. Therefore, MontiCore supports the declaration of editor-specific functionalities in a subsection of the DSLTool definition - the *concept texteditor*.

**Error Messages.** Adequate and understandable error messages are one crucial criterion for a successful adoption of a language. This is especially important for syntax-driven editors. Error messages should not be shown in textual form inside a console but should be integrated into the environment. The *problems view* inside Eclipse offers a possibility to show error messages and to connect them to a file and a line within that file. By selecting a marker the file is automatically opened and the line is highlighted.

MontiCore automatically produces *problem reports* for generated components which are shown inside the problems view inside Eclipse. Furthermore, hand-written components such as syntactic analyzers can be integrated into the parsing process by adding their class names in the *workflow* section of the editor concept. These components can produce problem reports which lead to the same effect as for generated components.

**Editor Actions.** It is possible to add context menu items to a generated text editor by implementing the interface shown in Figure 3 and adding the class name and the name of the context menu item in the editor concept. Since some functionalities such as refactoring depend on the specific position/string the user has selected, both information are forwarded to the hand-written code. Furthermore, the current editor is forwarded in order to gain access to the filename, to the whole project and thus to all other files within the project, or to write back refactored code.

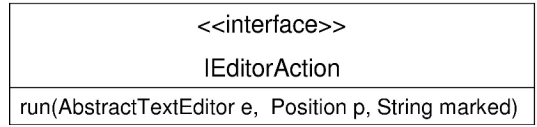

**Fig. 3.** Interface IEditorAction

Another prominent example for editor actions is code formatting. MontiCore allows to define pretty printers for languages on the fragment level which are combined automatically and can be included in an editor as *format action*.

**Navigator Actions.** Editor actions are hooked in the context menu of a generated text editor. However, there are often functionalities which depend on more than one input file, e.g. composition of models. Therefore, the developer can hook in hand-written functionalities into the *navigator view* of Eclipse similar to editor actions. This navigator view shows all projects and their contents such as files and folders within the workspace. The user is able to select different resources and to invoke the context menu depending on them. Again, the developer implements the interface as shown Figure 4 and adds the name of the implementing class and the name of the context menu item to the editor definition. For convenience, the `run` method has a parameter which maps each selected input file to its project.

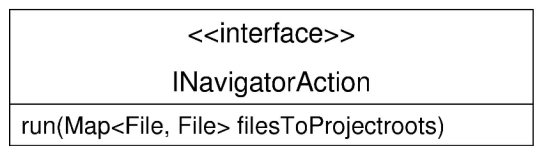

**Fig. 4.** Interface INavigatorAction

**Compositionality on Language Level.** In order to support compositional definitions of languages we decided to use a two-stage strategy. First, the language combination defined in the DSLTool is evaluated and all involved editor definitions are analyzed and combined. Second, we generate the functionalities as a plugin depending on the first stage resulting in a combined tool. Doing so, the functionalities such as the segments in the outline result from the combination of the definitions of all involved fragments.

The most important advantage is that the definitions are independent from each other regardless which fragments will be combined. Beyond that, there is no redundancy: each functionality is defined only once even if the fragment is used in several language combinations. Thus, changes in a fragment are immediately reflected in all language combinations (and therefore tools) the fragment is involved in.

We decided for a strict separation between language and fragment level functionalities as this enables to reuse fragments in a new composition as well as to reuse different languages in a new project without re-implementing the tool support. Thus, we are not only able to set up a library of composable language fragments but a library which consists of languages and language specific editors which can be reused in new project settings with minimal effort.

### 3.3 Compositionality on Tool Level

DSLs are typically defined to solve a clear cut task. Several languages may be used to describe all aspects of the system under design. In order to support the composition of languages and their specific tools into deployable products it is necessary to combine the editor definitions in a generated plugin instead of generating a plugin for all editors separately and to install them in parallel which is unpractical if tools shall handle cooperating languages. This is especially important if a fixed set of languages (e.g. a specific subset of the UML) is used to describe a system.

Therefore, MontiCore allows to define more than one editor in a single project. All grammars are handled separately and the plugin description/manifest file which is necessary for each Eclipse plugin is composed of all editor-specific information. However, it is still possible to generate plugins seperately for each language if desired and to install them in parallel. Furthermore, the generated parts can be supplemented by handwritten plugins such as a debugger or simulation engine.

## 4 A Demonstrating Example

To demonstrate the introduced capabilities of MontiCore we use a simplified and adapted version of message sequence charts [5]. MSCs are used to describe actions between objects or processes (referred to as *instances*). Instances transfer messages via *send* and *receive events*. Furthermore, *conditions* can be added to an instance, they are used to describe invariants in an appropriate language which have to be valid at a certain point of time. Conditions can be valid for a specific instance, at a bundle of instances, or all instances. Figure 5 shows a graphical and the according textual version of message sequence charts combined with Java for the conditions.

The first step towards a textual version is to define the grammar in MontiCore as outlined in Figure 6. MSCs consist of a name and set of `instances` and `Methods` (line 7). `Instances` in turn have also a name and several `Events` (line 9). `SendEvent`, `ReceiveEvent`, and `Condition` implement the `Event` interface and can therefore be used in an `instance` (line 11-21). Furthermore, `SendEvents` pass a `message` to a `receiver`, `ReceiveEvents` are defined analogously (line 11-15). `Conditions` (line 17-21) have a name and can be shared with all other instances (determined by the keywords `shared all`) or with specific instances (determined by `shared` followed by the the names of the instances in a comma-separated list). Both `Cond` and `Method` are marked as external rule (line 4 and 5) in order to combine the MSC language with any other language.

In order to define the properties of the generated editor on the fragment level the concept texteditor has to be added to the basic grammar as shown in lines 23-29 of Figure 6. The first part is to define the keywords which have to be colored in the code area in a comma separated list (line 24). Since we use the standard comments, the resulting editor automatically recognizes comment regions displayed in green.

*Foldable* code regions allow to collapse/uncollapse parts of the code in order to get more overview of the instance under design. A comma separated list (line 25) defines the nonterminals to be foldable, in this example an `MSC`, an `instance`, and a `condition`.

*Segments* describe the items to be shown inside the outline. As example we describe the segment for `SendEvents` (lines 26-27). The outline item should consist of a small icon (pict/arrow.gif) and an instance-specific text. This text in turn should be "Send to" followed by the name of the receiver, a colon, and the message.

In this example, MSCs should be combined with Java as action language, hence we have to define the Java grammar and the editor attributes in an analogous form to MSC. We omit details here for

```
         MontiCore-Grammar
1 msc mail{
2
3   instance sender{
4     out message to receiver;
5     in response from receiver;
6   }
7
8   instance receiver{
9     in message from sender;
10    condition inbox {
11       checkInbox()
12    }
13    out response to sender;
14  }
15
16  public boolean checkInbox() {
17    return receiver.messages > 0;
18  }
19 }
```

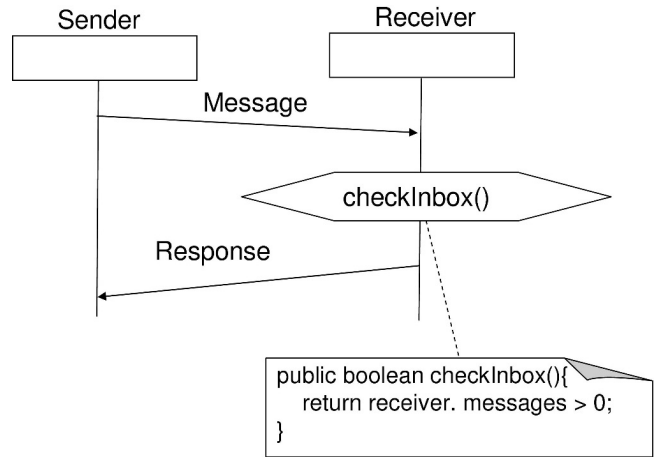

**Fig. 5.** Simple MSCs in diagrammatical and textual version

the sake of space. Note that both definitions are completely standalone, i.e. the MSC grammar and the corresponding concept editor attributes do not know anything about the grammar/editor they are combined with.

In order to combine the editors on the language level, the concept texteditor has to be added to the DSLTool definition as shown in Figure 7. Lines 3-5 combine the languages: the MSC-parser is used as start-parser (line 2), Java expressions are embedded into MSCs by replacing the external nonterminal `cond` (line 3) whereas methods are used to replace the external nonterminal `method` (line 4/5). Lines 7-10 combines the pretty printers which leads to a *format action* in the resulting editor as described in Section 3.2.

The concept texteditor consists of several parts which describe the behavior and the functionalities of the resulting editor. We describe the most important points. Line 13 defines the class of the tool which is used to process the text inside the editor. This tool is manually written, command line arguments can be used in order to define the *workflows* which have to be executed (parsing, symbol table building, etc.). Line 14 defines additional *workflows* which should be executed when the editor evaluates the text. We implemented 2 workflows: `symtab` for building a symbol table and `check` for checking context conditions. The declaration of `symtab` and `check` leads to arguments which are passed to the tool. These workflows are able to add *problem reports* to the processed file leading to error messages inside Eclipse (c.f. Section 3.2).

The last functionalities of our example are *editor* and *navigator actions*. The former are added to the context menu of the generated editor, the latter can be used inside the context menu of the navigator. Both actions consist of a name to be shown in the context menu and the name of class which realizes the functionality. As example we implemented an algorithm which generates traces for an MSC which is realized as *editor action* in line 15. Furthermore, we implemented a class which vertically composes 2 or more MSCs as *navigator action* (line 17).

```
                                               MontiCore-Grammar
1 grammar MSC {
2
3   interface Event;
4   external Cond;
5   external Method;
6
7   MSC = "msc" name:IDENT "{" (Instance | Method)* "}";
8
9   Instance = "instance" name:IDENT "{" Event* "}";
10
11  SendEvent implements Event =
12    "out" message:IDENT "to" receiver:IDENT ";";
13
14  ReceiveEvent implements Event =
15    "in" message:IDENT "from" sender:IDENT ";";
16
17  Condition implements Event = "condition" name:IDENT
18    (shared:["shared"]
19      (sharedWithAll:["all"] | sharedWith:IDENT ("," sharedWith:IDENT)*)
20    )?
21    ( "{" Cond "}" | ";");
22
23  concept texteditor {
24    keywords: msc, instance, in, out, to, from, condition, shared, all;
25    foldable: MSC, Instance, Condition;
26    segment: SendEvent ("pict/arrow.gif")
27      show: "Send to " receiver ":" message;
28    // further segments
29  }
30 }
```

**Fig. 6.** Definition of MSCs

The definitions desribed so far are sufficient to generate a language specific editor for MSCs combined with Java as embedded language. The most important functionalities are shown in Figure 8 and Figure 9. Furthermore, Table 1 shows some key figures of the example.

|  | Lines of code |
|---|---:|
| DSL-specification (handwritten) | 2,379 |
| Java-Code (handwritten code generation) | 3,296 |
| Java-Code (generated editor) | 2,413 |

**Table 1.** Key figures for MSC example

```
 1 rootfactory MSCRootFactory for MSCRoot<MCCompilationUnit>{
 2   mc.examples.msc.msc.MSC.MCCompilationUnit mscdefinition <<start>>;
 3   mc.examples.msc.java.JavaDSL.Expression cond in mscdefinition.cond;
 4   mc.examples.msc.java.JavaDSL.MethodDeclaration method in
 5     mscdefinition.method;
 6
 7   prettyprint {
 8     mc.examples.msc.msc.prettyprint.MSCConcretePrettyPrinter;
 9     mc.examples.msc.java.JavaDSLConcretePrettyPrinter;
10   }
11 }
12 concept texteditor {
13   tool: "mc.examples.msc.msc.MSCTool";
14   workflows: symtab, check;
15   menuitem Generate Trace
16     ("mc.examples.msc.msc.action.GenerateTraceAction");
17   navigatoritem Compose
18     ("mc.examples.msc.msc.compose.ComposeAction");
19 }
20 ...
```

**Fig. 7.** Excerpt from language and editor composition for MSCs and Java

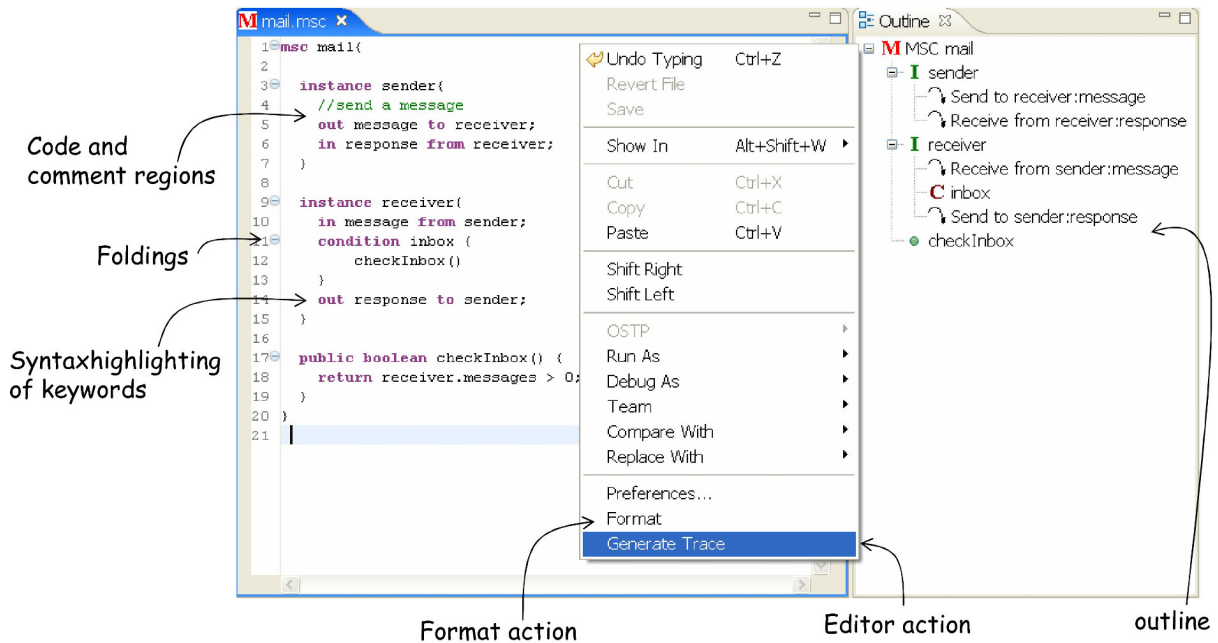

**Fig. 8.** Resulting Editor for MSCs

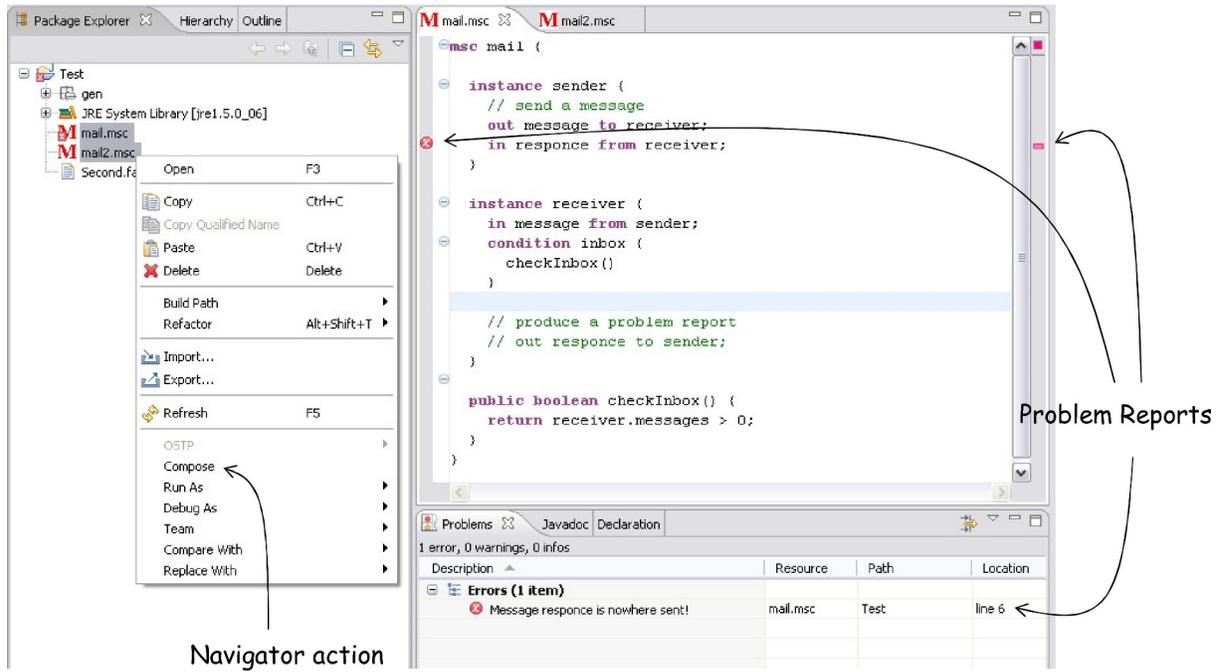

**Fig. 9.** Problem Reports and Navigator Actions

## 5 Related work

DSL are often developed using special language workbenches [11, 13, 17] that supply the user with help to develop the language efficiently. These workbenches allow the definition of languages and the generation of editors that look similar to the meta environment as they use a common code base. In contrast to our approach [11] and [13] generate graphical editors.

The SAFARI project [16] allows to develop programming languages and their Eclipse integration. In contrast to their approach, we focused on a pure textual approach that is needless integrated with our language definition and avoided wizards that - to our impression - hinder language evolution.

In [6] is described how textual syntax and as a further step the frontend of a compiler can be derived from an abstract syntax as first step towards a generation of text-based IDE. The generation of language-specific editors is not performed.

XText is part of the openArchitectureWare project [15] and can be used in order to generate Eclipse-based editors from a grammar definition. These editors provide similar functionalities to the editors generated by MontiCore. However, there are some important differences between xText and our approach. The most important is that we aim at compositional DSL and tools, wheras xText concentrates on standalone grammars.

## 6 Conclusion

In this paper we introduced an extension of the MontiCore grammar format to generate Eclipse plugins for simplifying the development of DSLs. We especially investigated how modular language development can supported by modular editor definitions. We identified elements that are best defined at fragment, language or tool level to maximize the reuse for future development.

In the future we like to explore the generation of further tool support for Eclipse by supplying the language developer with a convenient interface for defining refactorings for the language under design. In addition graphical outlines that supplement the textual interface could be generated from the user input. As proof of concept we are currently bootstrapping an editor for MontiCore itself.

**Acknowledgement:** The work presented in this paper is undertaken as a part of the MODELPLEX project. MODELPLEX is a project co-funded by the European Commission under the "Information Society Technologies" Sixth Framework Programme (2002-2006). Information included in this document reflects only the authors' views. The European Community is not liable for any use that may be made of the information contained herein.